\documentclass[amsmath,amssymb,twocolumn,prl,superscriptaddress]{revtex4}
\usepackage{amsfonts}
\usepackage{amsmath}
\usepackage{amsthm}
\usepackage{amscd}
\usepackage{amssymb}
\usepackage{subfigure}
\usepackage{amsxtra}
\usepackage{gensymb}
\usepackage{bm}           
\usepackage{bbm}
\usepackage{graphicx}
\usepackage{epstopdf}
\usepackage{color}
\usepackage{times}
\usepackage{mdframed}
\usepackage[colorlinks=true,citecolor=blue,urlcolor=black]{hyperref}

\def\bi#1\ei {\begin{itemize}#1\end{itemize}}
\def\bn#1\en {\begin{enumerate}#1\end{enumerate}}
\def\bea#1\eea {\begin{align}#1\end{align}}
\def\bean#1\eean {\begin{align*}#1\end{align*}}
\def\ben#1\een {\begin{equation*}#1\end{equation*}}
\def\be#1\ee {\begin{equation}#1\end{equation}}
\def\bes#1\ees {\begin{equation}\begin{split}#1\end{split}\end{equation}}
\def\bear#1\eear {\begin{eqnarray}#1\end{eqnarray}}
\def\bear#1\eear {\begin{eqnarray*}#1\end{eqnarray*}}

\newcommand{\beq}{\begin{equation}}
\newcommand{\eeq}{\end{equation}}
\newcommand{\bra}[1]{\ensuremath{\langle#1|}}
\newcommand{\ket}[1]{\ensuremath{\left|#1\right\rangle}}

\begin{document}

\title{\bf Experimental Quantum Digital Signature over 102 km}

\author{Hua-Lei Yin}
\author{Yao Fu}
\author{Hui Liu}
\author{Qi-Jie Tang}
\author{Jian Wang}
\affiliation{National Laboratory for Physical Sciences at Microscale and Department of Modern Physics, University of Science and Technology of China, Hefei, Anhui 230026, China}
\affiliation{The CAS Center for Excellence in QIQP and the Synergetic Innovation Center for
QIQP, University of Science and Technology of China, Hefei, Anhui 230026, China}

\author{Li-Xing You}
\author{Wei-Jun Zhang}
\author{Si-Jing Chen}
\author{Zhen Wang}
\affiliation{State Key Laboratory of Functional Materials for Informatics, Shanghai Institute of Microsystem and Information Technology, Chinese Academy of Sciences, Shanghai 200050, China}

\author{Qiang Zhang}
\email{qiangzh@ustc.edu.cn}
\affiliation{National Laboratory for Physical Sciences at Microscale and Department of Modern Physics, University of Science and Technology of China, Hefei, Anhui 230026, China}
\affiliation{The CAS Center for Excellence in QIQP and the Synergetic Innovation Center for
QIQP, University of Science and Technology of China, Hefei, Anhui 230026, China}

\author{Teng-Yun Chen}
\email{tychen@ustc.edu.cn}

\author{Zeng-Bing Chen}
\email{zbchen@ustc.edu.cn}

\author{Jian-Wei Pan}
\email{pan@ustc.edu.cn}
\affiliation{National Laboratory for Physical Sciences at Microscale and Department of Modern Physics, University of Science and Technology of China, Hefei, Anhui 230026, China}
\affiliation{The CAS Center for Excellence in QIQP and the Synergetic Innovation Center for
QIQP, University of Science and Technology of China, Hefei, Anhui 230026, China}

\date{\today}

\begin{abstract}
Quantum digital signature (QDS) is an approach to guarantee the nonrepudiation, unforgeability and transferability of a signature with the information-theoretical security. All previous experimental realizations of QDS relied on an unrealistic assumption of secure channels and the longest distance is only several kilometers. Here, we have experimentally demonstrated a recently proposed QDS protocol without any secure channel. Exploiting the decoy state modulation, we have successfully signed one bit message through up to 102 km optical fiber. Furthermore, we continuously run the system to sign the longer message ``USTC" with 32 bit at the distance of 51 km.  Our results pave the way towards the practical application of QDS.
\end{abstract}

\maketitle

Digital signature \cite{diffie:1976:new} is a basic primitive for plenty of cryptographic protocols, which has many applications in software distribution, financial transactions, contract management software and so on. Classical digital signature mainly exploits the Rivest-Shamir-Adleman (RSA) protocol \cite{rivest:1978:method}, whose security is based on the mathematical complexity of integer factorization problem. This, however, may become vulnerable with a quantum computer \cite{shor:1994:algorithms}. By exploiting the laws of quantum mechanics, quantum key distribution (QKD) can offer two legitimate users to share the random key with information-theoretical security \cite{Gisin:2002:Quantum,Scarani:2009:The}. Similarly, one can expect to exploit the laws of quantum mechanics to sign a message with the information-theoretical security, which is called quantum digital signature (QDS).

A basic digital signature model will introduce at least three authorized parties, in addition, the three authorized parties cannot be assumed all honest. By contrast, a conventional QKD system has two authorized and honest parties. This is why QKD has entered practical application and networking deployment \cite{qiu:2014:quantum}, while the QDS is still on the stage of the security analysis and the proof-of-principle experimental demonstration.

The first QDS protocol was proposed by Gottesman and Chuang in 2001 \cite{gottesman:2001:quantum}, where several technical challenges need to be fixed for a practical implementation, including nondestructive state comparison, long time quantum memory and secure quantum channel. Thereafter, QDS has attracted a great deal of interest in the literature. Various QDS protocols have been proposed \cite{Andersson:2007:Experimentally,Dunjko:2014:Quantum,Arrazola:2014:Quantum,Wallden:2015:Quantum,arrazola2015multiparty} and some pioneering experimental efforts have been made towards this direction \cite{clarke2012experimental,Collins:2014:Realization,Donaldson:2016:Experimental,croal2016free}. To name a few, Clarke \emph{et al}. \cite{clarke2012experimental} utilize coherent states and linear optics to avoid the nondestructive operation and provides the first experimental try. Collins \emph{et al}. \cite{Collins:2014:Realization} present a realization without the need of quantum memory, which, however, still needs the assumption of secure quantum channel. Secure quantum channel means that the quantum channel should not be tampered. Note that the basic model of quantum communication such as QKD \cite{Gisin:2002:Quantum,Scarani:2009:The} and quantum secret sharing \cite{hillery:1999:quantum} is that the quantum channel can be eavesdropped and tampered with. Therefore, the secure quantum channel is an unrealistic assumption and limits the application of QDS. Meanwhile, the Mach-Zehnder interferometer configuration in the experiment by Collins \emph{et al}. \cite{Collins:2014:Realization} requires phase stability between distant parties, which is experimentally challenging for a long distance implementation.

Very recently, new QDS protocols \cite{Yin:2016:Practical,Amiri:2016:Secure} have been proposed to remove the assumption of secure quantum channel. A kilometer range demonstration for the new protocols is provided \cite{Donaldson:2016:Experimental}, which however introduces another assumption of secure classical channel. In this Letter, we provide a complete QDS experiment without quantum or classical secure channel assumption over 102 km optical fiber. We do believe that with these experimental advances, QDS with information-theoretical security will come to practical applications soon.

\begin{figure*}[tbh]
\centering
\resizebox{14cm}{!}{\includegraphics{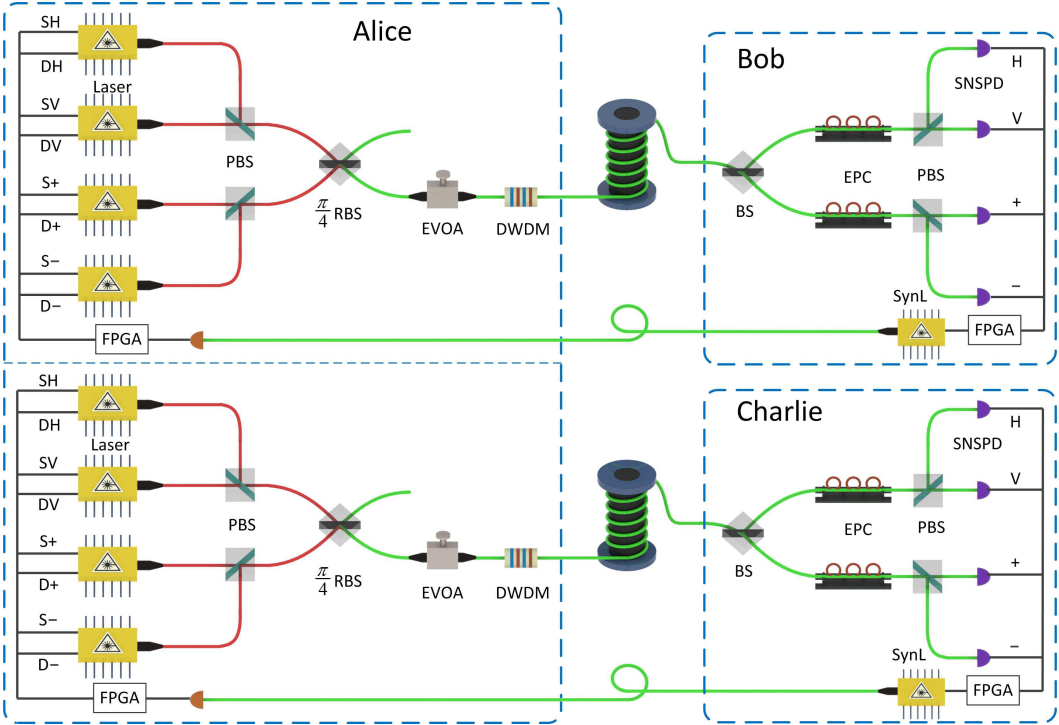}}
\caption{Experimental setup for quantum digital signature. Alice randomly prepares two copies of BB84 states with decoy-state method and sends to Bob and Charlie through two fiber spools, respectively. Bob and Charlie detect the photon with their SNSPDs (superconducting nanowire single-photon detector). PBS: polarization beam splitter, $\frac{\pi}{4}$RBS: $\pi/4$ rotation beam splitter, EVOA: electrical variable optical attenuator, DWDM: dense wavelength division multiplexer, BS: beam splitter, EPC: electric polarization controller, FPGA: filed programmable gate array, SynL: synchronization laser.}
\label{f1}
\end{figure*}

Before describing the experiment in detail, we first introduce the QDS protocol \cite{Yin:2016:Practical} used in this work. In a digital signature protocol, Alice, the sender, will send a message with a digital signature to two recipients, Bob and Charlie. Without loss of generality, we take Bob as the authenticator. He then forwards the information that he received from Alice, to Charlie. In a successful digital signature protocol, Alice could not deny the signature, which is called nonrepudiation. On the other hand, Bob could not forge the message, which is called unforgeability.
If Bob accepts the message, Charlie will also accepts the message, which is called transferability.

Our protocol is divided into two steps, quantum stage and signature stage. In the quantum stage, for each future possible message $m=0,1$, Alice exploits weak coherent states (WCS) to randomly prepare two identical qubit states from the BB84 states \cite{bennett1984quantum}, $\ket{H}$, $\ket{V}$, $\ket{+}$ and $\ket{-}$, where  $\ket{H}$, $\ket{V}$ represent horizontal and vertical polarization states,  $\ket{+}=(1/\sqrt{2})(\ket{H}+\ket{V})$ and $\ket{-}=(1/\sqrt{2})(\ket{H}-\ket{V})$. In order to avoid photon-number splitting attack, Alice exploits the decoy-state method by randomly varying the intensity of the pulses. She chooses three intensities $\mu,\nu,\omega$, one as signal and two as decoy states. Then, Alice randomly sends one qubit state with intensity of $\alpha$ to Bob and the other with intensity of $\beta$ to Charlie, where, $(\alpha,\beta) \in (\mu,\nu,\omega)$. Note that, the polarization states for Bob and Charlie are identical, while the intensities are not necessary to be the same.

Bob and Charlie independently and randomly exploit $Z$ or $X$ basis to measure the received quantum state. Alice, Bob and Charlie record the corresponding data when both Bob's and Charlie's detectors have a click. The nonorthogonal state encoding scheme \cite{Scarani:2004:Quantum} are used to identify the conclusive outcome and inconclusive outcome. For each quantum state, Bob (Charlie) compares his measurement outcomes with two nonorthogonal states announced by Alice. If his measurement outcome is orthogonal to one of Alice's announced states, he concludes a conclusive result that the other state has been sent. Otherwise, he concludes that it is an inconclusive outcome, which is only known by himself.

Then, the signature process starts. Alice announces the nine intensity sets and also the bit information of six intensity sets, $\mu\omega$, $\omega\mu$, $\nu\nu$, $\nu\omega$, $\omega\nu$ and $\omega\omega$. Charlie, as the verifier, estimates the yield, $Y_{11}^{C}$ and the quantum bit error rate $e_{11}^{C}$ for single-photon pairs of his conclusive results \cite{Yin:2016:Practical}, where a single-photon pair represents that one photon is sent to Bob and one photon is sent to Charlie.
Exploiting the entanglement distillation technique \cite{Yin:2016:Practical,Tamaki:2006:Unconditionally,Yin:2016:Security},
the min-entropy of Bob about Charlie's conclusive results with the single-photon pairs can be bounded by $1-H(e_{p11}^{C}|e_{11}^{C})$, where $e_{p11}^{C}$ is the phase error rate and $H(e_{p11}^{C}|e_{11}^{C})$ is the conditional Shannon entropy (see the Supplemental Material for details). Given the bound of the min-entropy of Bob, one can acquire the lower bound of mismatching rate $S_{11}$ between Bob's declaration and Charlie's conclusive results with single-photon pairs, which can be given by \cite{Yin:2016:Practical,Amiri:2016:Secure},
\begin{equation} \label{eq1}
\begin{aligned}
1-H(e_{p11}^{C}|e_{11}^{C})-H(S_{11})=0,
\end{aligned}
\end{equation}
where $H(x)=-x\log_{2}x-(1-x)\log_{2}(1-x)$ is the Shannon entropy function. Exploiting the mismatching rate $S_{11}$, one can restrict the forgery attack of Bob.

\begin{table*}
\begin{ruledtabular}
\caption{\label{tab:Key} The error rates and secure thresholds at different distances in our experiment. Thereinto,  $\varepsilon_{\textrm{rep}}$ ($\varepsilon_{\textrm{for}}$) represents the probability of successful repudiation (forgery) attack.  $N$ represents the total pulse pairs sent by Alice to sign half bit.}
\begin{tabular}{c|cccccccc}
Distance
&
\multicolumn{2}{c}{25 km}
&
\multicolumn{2}{c}{51 km}
&
\multicolumn{2}{c}{76 km}
&
\multicolumn{2}{c}{102 km}\\
Attenuation
&
\multicolumn{2}{c}{4.9 dB}
&
\multicolumn{2}{c}{9.8 dB}
&
\multicolumn{2}{c}{14.8 dB}
&
\multicolumn{2}{c}{19.8 dB}\\

$T_{v}$
&
\multicolumn{2}{c}{2.0\%}
&
\multicolumn{2}{c}{2.0\%}
&
\multicolumn{2}{c}{1.9\%}
&
\multicolumn{2}{c}{2.2\%}\\

$T_{a}$
&
\multicolumn{2}{c}{0.6\%}
&
\multicolumn{2}{c}{0.6\%}
&
\multicolumn{2}{c}{0.55\%}
&
\multicolumn{2}{c}{0.7\%}\\
\hline
Message
& m=0 & m=1 & m=0 & m=1 & m=0 & m=1 & m=0 & m=1\\
$E_{s}^{B}$
& 0.35\%& 0.39\% & 0.37\% & 0.37\% & 0.36\% & 0.29\% & 0.51\% & 0.45\%\\
$E_{s}^{C}$
& 0.26\%& 0.29\% & 0.25\% & 0.22\% & 0.30\% & 0.26\% & 0.42\% & 0.40\%\\
$S_{11}$
& 4.33\%& 4.21\% & 4.27\% & 4.35\% & 4.28\% & 4.10\% & 4.46\% & 4.42\%\\
Time
& 20s & 20s & 180s & 180s & 1620s & 1620s & 33420s & 33420s \\
$N$
& $1.5\times10^9$ & $1.5\times10^9$ & $1.35\times10^{10}$ & $1.35\times10^{10}$ & $1.215\times10^{11}$ & $1.215\times10^{11}$ & $2.5065\times10^{12}$ & $2.5065\times10^{12}$\\
$\varepsilon_{\textrm{rep}}$
& $7.1\times10^{-10}$ & $1.4\times10^{-6}$ & $1.6\times10^{-10}$ & $3.4\times10^{-11}$ & $5.3\times10^{-8}$ & $4.9\times10^{-12}$ & $4.9\times10^{-8}$ & $5.7\times10^{-13}$\\
$\varepsilon_{\textrm{for}}$
& $7.4\times10^{-15}$ & $5.6\times10^{-9}$ & $4.1\times10^{-13}$ & $4.1\times10^{-12}$ & $2.5\times10^{-8}$ & $3.7\times10^{-9}$ & $1.4\times10^{-19}$ & $7.0\times10^{-10}$\\
$\varepsilon_{\textrm{rob}}$
& $8.2\times10^{-12}$ & $3.9\times10^{-8}$ & $1.6\times10^{-9}$ & $4.6\times10^{-10}$ & $1.2\times10^{-7}$ & $1.2\times10^{-14}$ & $2.2\times10^{-9}$ & $2.0\times10^{-18}$\\
\end{tabular}
\end{ruledtabular}
\end{table*}

On the other hand, the data string under the case of three intensity sets $\mu\mu$, $\mu\nu$, $\nu\mu$ constitute an overall data string. Charlie randomly chooses some data from the overall data string as the sampling data string and informs to Alice and Bob. They compare the bit value for the sampling string and  estimate the quantum bit error rate of conclusive results, $E_{s}^{B}$ and $E_{s}^{C}$, which are utilized to restrict repudiation from Alice. The remaining data string of Alice, Bob and Charlie are kept for the digital signature, denoted as $S_{Am}$, $S_{Bm}$ and $S_{Cm}$, respectively.

From the view of Alice, the status of all data (conclusive results and inconclusive results) owned by Bob (Charlie) could be regarded as the same, since Bob (Charlie) do not announce the position of conclusive result. By random sampling, the upper bound of the difference between the data owned by Bob and Charlie can be bounded. With this restrict, Alice, has to send almost the same quantum states to Bob and Charlie and thus the potential repudiating attack is avoided. Taking into account the finite-size effect \cite{Ma:2012:Statistical,korzh2015provably,chernoff1952measure,curty2014finite}, the authentication (verification) security threshold $T_{a}$ ($T_{v}$) can be determined.  With $T_{a}$ and $T_{v}$, one can calculate the probabilities of successful repudiation attack, forgery attack and the robustness. Detailed analysis can be found in the Supplemental Material.

To sign one-bit message $m$, Alice sends the message and the corresponding data string $(m,S_{Am})$ to the authenticator Bob. Bob will accept the message when the mismatching rate of his conclusive outcome is less than $T_{a}$. If Bob accepts the message, he forwards $(m,S_{Am})$ to the verifier, Charlie. Charlie will accept the message when the mismatching rate of his conclusive outcome is less than $T_{v}$.

In the implementation, the quantum stage setup is shown in Fig. \ref{f1}. Alice prepares four polarization-encoded BB84 states with four electrically modulated distributed feedback laser diodes. The emissions of the laser diodes are centered at 1550 nm with a pulse duration of 0.4 ns and repetition frequency of 75 MHz. The difference of the central wavelength from these lasers is well controlled to be less than 0.02 nm via temperature control. We combine the four laser diodes with two PBS and one 45 degree RBS into a single fiber. An electrical variable optical attenuator (EVOA) is used to attenuate the average photon number per pulse to the experimental level. The dense wavelength division multiplexer (DWDM) with 100 GHz bandwidth is used to filter any spurious emission. After the filtration, the quantum states are sent out to Bob through a fiber spool.

We exploit the decoy-state method \cite{Hwang:2003:Decoy,Wang:2005:Beating,Lo:2005:Decoy} by varying the injection electrical current for the laser diodes. We set the intensities of signal states $\mu=0.22$, decoy states $\nu=0.066$ and vacuum states $\omega=0$ and their corresponding probability distributions are $P_{\mu}=65\%$, $P_{\nu}=35\%$ and $P_{\omega}=5\%$, respectively. All random signals for choosing polarization states or intensities are all derived from random numbers generated beforehand. Meanwhile, the phases for the directly modulated laser diode are random, which is immune to the unambiguous state discrimination attack \cite{Tang:2013:USD}.

In Bob's side, the detector system contains four superconducting nanowire single-photon detectors (SNSPD) that provide the detection efficiency of 52\% at the dark count rate of 10 counts per second. A polarization measurement module is connected to the detector system via single-mode fibers and consists of one beam splitter (BS), two electric polarization controllers (EPC) and two PBS. The EPCs are used for compensation of the polarization fluctuation in the fiber spool. The optical pulses go through the polarization measurement module to be detected by the SNSPD. The insertion loss of the polarization measurement module is around 1.2 dB.

Bob exploits a crystal oscillator circuit to generate 500 kHz electric signals as the synchronization signals of system. Bob sends synchronization laser pulses (SLP) at 1570 nm modulated by the 500 KHz electric signals to Alice through an additional fiber. A photoelectric detector (PD) and phase-locked loop utilized by Alice detect the SLP and regenerate a system clock frequency of 75 MHz by frequency multiplication as the clock for her four laser diodes. Alice exploits the same setup to send quantum states to Charlie.

In our experiment, we perform a symmetrical case that each fiber length from Alice to Bob and Alice to Charlie are almost the same. The length of fiber spool are 25 km, 51 km, 76 km and 102 km, respectively. For each distance, we send two groups of quantum states to sign one future bit message in the signature stage, where the first group is used to sign future message bit $m=0$ and the second group for bit $m=1$. All the parameter estimation and the message signature are implemented in a local area network connecting the three users.

The bit error rates $E_{s}^{B}$ and $E_{s}^{C}$ of Bob's and Charlie's conclusive results in the sampling data string are listed in Table \ref{tab:Key}. Exploiting the decoy-state method \cite{Yin:2016:Practical}, the yield and quantum bit error rate of single-photon pairs can be acquired. The lower bound of the mismatching rate $S_{11}$ can be calculated using Eq. \eqref{eq1} which is shown in Table \ref{tab:Key}. Given that  the security bound is $\varepsilon_{\textrm{sec}}<10^{-5}$ and the robustness bound is $\varepsilon_{\textrm{rob}}<10^{-6}$, the authentication and verification security thresholds $T_{a}$ and $T_{v}$  can be chosen with proper values, which are also shown in Table \ref{tab:Key}. More details of experimental results can be found in the Supplemental Material.

\begin{figure}[tbh]
\centering
\resizebox{8.5cm}{!}{\includegraphics{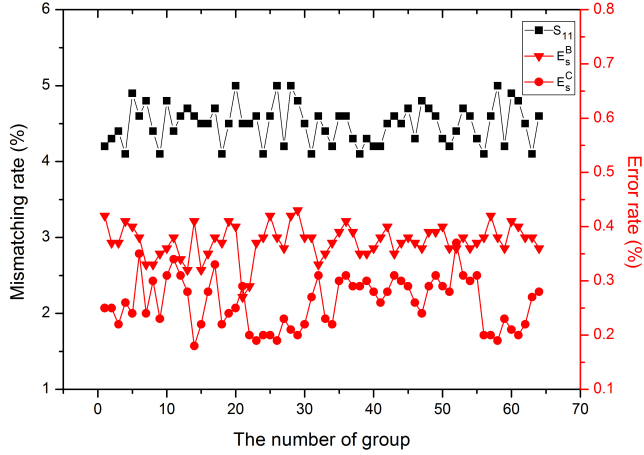}}
\caption{The error rates and the mismatching rates for each group. The experimental error rates $E_{s}^{B}$ ($E_{s}^{C}$) of Bob's (Charlie's) conclusive results in the sampling data string are almost  0.3\%--0.4\% (0.2\%--0.3\%). The mismatching rate $S_{11}$ calculated by Eq. \eqref{eq1} are almost 4.2\%--5.0\%}
\label{f2}
\end{figure}

\begin{figure}[tbh]
\centering
\resizebox{8cm}{!}{\includegraphics{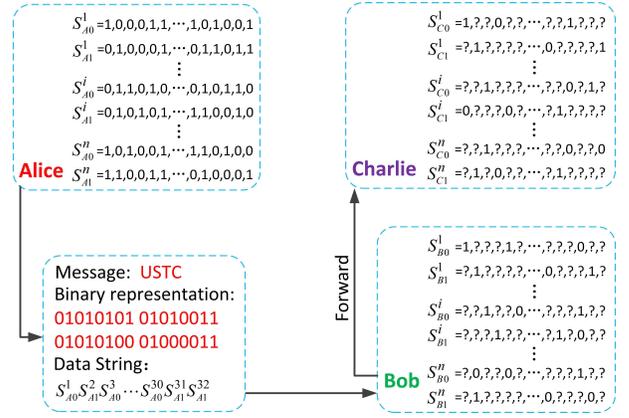}}
\caption{Demonstration of signing the message string ``USTC". Alice sends the ASCII code for the message ``01010101010100110101010001000011" and the corresponding data string $S_{A0}^{1}S_{A1}^{2}\cdots S_{A1}^{31}S_{A1}^{32}$ to Bob through the authenticated classical channel.
Bob compares the data string $S_{A0}^{1}S_{A1}^{2}\cdots S_{A1}^{31}S_{A1}^{32}$ and $S_{B0}^{1}S_{B1}^{2}\cdots S_{B1}^{31}S_{B1}^{32}$, and  accepts the message since the error rates of Bob's conclusive rates is less than $T_{a}$ for each group. Bob forwards the message and the corresponding data string to Charlie through the authenticated classical channel. Charlie compares the data string $S_{A0}^{1}S_{A1}^{2}\cdots S_{A1}^{31}S_{A1}^{32}$ and $S_{C0}^{1}S_{C1}^{2}\cdots S_{C1}^{31}S_{C1}^{32}$, and  accepts the message since the error rates of Charlie's conclusive rates is less than $T_{v}$ for each group.
}
\label{f3}
\end{figure}

Except for proof-of-principle demonstration of a one-bit QDS like all previous experimental demonstrations, we also implement QDS for a longer message. We continuously collect 64 groups and each group has 180 seconds at the distance of 51 km.  The bit error rates of Bob's  and Charlie's conclusive results $E_{s}^{B}$ and $E_{s}^{C}$ in the sampling data string for each group are shown in Fig. \ref{f2}. We set the security bound to be $\varepsilon_{\textrm{sec}}<10^{-5}$ and the robustness bound to be $\varepsilon_{\textrm{rob}}<10^{-6}$ for each group. For simplicity, the authentication and verification security thresholds $T_{a}$ and $T_{v}$  can be fixed to be 2.0\% and 0.6\%, respectively. Note that the secure thresholds can be different for each group.  We can sign 32 bit message since two groups need to be used to sign one bit message. Before Alice signs the long message, she will publicly announce the length of the message bits. Here, we have successfully signed a 32-bit message ``USTC". The process of signing the message can be found in Fig. \ref{f3}.

In summary, we have experimentally demonstrated a QDS protocol without the assumption of any secure channel. Exploiting the decoy state modulation and the BB84 state encoding, we have successfully signed one bit message through up to 102 km optical fiber. Furthermore, we continuously run the system to sign the longer message ``USTC" with 32 bits at the distance of 51 km.  We remark that it needs 360 seconds to sign one bit message at the distance of 51 km, which currently seems to be not so practical. However, if we implement the full parameter optimization and joint constrained statistical fluctuation \cite{wang:2016:making}, combined with the six-state encoding \cite{Yin:2016:Practical}, the signature rate will increase obviously with more than two orders of magnitude.

This work has been supported by the National Fundamental Research Program (under Grant No. 2013CB336800), the National Natural Science Foundation of China (under Grant No. 61125502), the Chinese Academy of Science, the 10000-Plan of Shandong Province and the Science Fund of Anhui Province for Outstanding Youth.

\newpage

\onecolumngrid

\section{Decoy state scheme and finite-size effect}
In this section, we will review the probability of repudiation and forgery attack calculations for the quantum digital signature (QDS) protocol.
No one can unambiguously discriminate two copies of quantum states from the four polarization states $\ket{H},\ket{V},\ket{+},\ket{-}$. For the two-photon components, the min-entropy of Bob about Charlie's conclusive results acquired by the nonorthogonal state encoding scheme \cite{Scarani:2004:Quantum} can be quantified by the entanglement distillation technique \cite{Yin:2016:Practical,Tamaki:2006:Unconditionally,Yin:2016:Security}. The relationship between phase error rate $e_{p}$ and the bit error rate $e_{b}$ is given by
\begin{equation} \label{eq1}
\begin{aligned}
e_{p}&=\min_{x}\left\{xe_{b}+f(x)\right\}, \forall x, ~~\frac{2-\sqrt{2}}{4}e_{b}\leq a\leq\frac{2+\sqrt{2}}{4}e_{b},
\end{aligned}
\end{equation}
and
\begin{equation} \label{eq2}
\begin{aligned}
f(x)=\frac{3-2x+\sqrt{6-6\sqrt{2}x+4x^2}}{6},
\end{aligned}
\end{equation}
where $a$ is the probability that both bit flip and phase shift occur, which quantifies the mutual information between phase and bit errors. The conditional Shannon entropy function can be given by \cite{Yin:2016:Practical}
\begin{equation} \label{mutual information}
\begin{aligned}
H(e_{p}|e_{b})=&-(1+a-e_{b}-e_{p})\log_{2}\frac{1+a-e_{b}-e_{p}}{1-e_{b}}-(e_{p}-a)\log_{2}\frac{e_{p}-a}{1-e_{b}}-(e_{b}-a)\log_{2}\frac{e_{b}-a}{e_{b}}-a\log_{2}\frac{a}{e_{b}}.
\end{aligned}
\end{equation}

The intensity set $\alpha\beta$ represents that Alice sends weak coherent state pulses to Bob with intensity $\alpha$ and weak coherent state pulses to Charlie with intensity $\beta$.
Alice prepares the phase randomized weak coherent state pulse pairs in $Z$ basis or $X$ basis with the intensity sets of $\mu\mu$, $\mu\nu$, $\nu\mu$, $\mu0$, $0\mu$, $\nu\nu$, $\nu0$, $0\nu$, $00$. In the photon number space, the density matrix for a pulse pair of intensity $\alpha\beta$ can be given by
\begin{equation} \label{gain}
\begin{aligned}
\rho_{\alpha\beta}=\sum_{n=0}^{\infty}\sum_{m=0}^{\infty}e^{-\alpha}\frac{\alpha^{n}}{n!}e^{-\beta}\frac{\beta^{m}}{m!}\ket{n}\bra{n}\ket{m}\bra{m}.
\end{aligned}
\end{equation}
The effective detection event can be defined as that both Bob and Charlie have a detection click.
We denote $N_{\alpha\beta}$ as the number of pulses sent by Alice with the intensity set $\alpha\beta$. $M_{\alpha\beta}$ is the number of effective detection events.
$M_{\alpha\beta}^{B}$ ($M_{\alpha\beta}^{C}$) is the number of effective detection events given that Bob (Charlie) has the conclusive results.
The gain $Q_{\alpha\beta}^{C}$ is the ratio of $M_{\alpha\beta}^{C}$ to $N_{\alpha\beta}$. $E_{\alpha\beta}^{C}$ is the quantum bit error rate in $M_{\alpha\beta}^{C}$ events.

Denote $Y_{11}^{C}$ ($e_{11}^{C}$) as the yield (bit error rate) of Alice sending single-photon pairs and Charlie having conclusive results.
Exploiting the decoy-state method \cite{Yin:2016:Practical}, the lower bound of yield $Y_{11}^{C}$ and the upper bound of $e_{11}^{C}$ with analytic form can be written as
\begin{equation} \label{yield3}
\begin{aligned}
Y_{11}^{C}\geq & \frac{1}{\mu^{2}\nu^{2}(\mu-\nu)}\times\Big\{\mu^{3}\big[e^{2\nu}Q_{\nu\nu}^{C}-e^{\nu}\big(Q_{\nu0}^{C}+Q_{0\nu}^{C}\big)\big]-\nu^{3}\big[e^{2\mu}Q_{\mu\mu}^{C}-e^{\mu}\big(Q_{\mu0}^{C}+Q_{0\mu}^{C}\big)\big]+(\mu^{3}-\nu^{3})Q_{00}^{C}\Big\}\\
\end{aligned}
\end{equation}
and
\begin{equation} \label{error3}
\begin{aligned}
e_{11}^{C}\leq & \frac{1}{\nu^{2}Y_{11}^{C}}\Big[e^{2\nu}E_{\nu\nu}^{C}Q_{\nu\nu}^{C}-e^{\nu}\big(E_{\nu0}^{C}Q_{\nu0}^{C}+E_{0\nu}^{C}Q_{0\nu}^{C}\big)+E_{00}^{C}Q_{00}^{C}\Big].
\end{aligned}
\end{equation}
We exploit the standard error analysis method \cite{Ma:2012:Statistical} to calculate the finite-size effect of decoy state estimation.
Thus, we have
\begin{equation} \label{sf}
\begin{aligned}
Q_{\mu\nu}^{CU}=Q_{\mu\nu}^{C}\left(1+ \frac{\gamma}{\sqrt{N_{\mu\nu}Q_{\mu\nu}^{C}}}\right),~~~~~~~Q_{\mu\nu}^{CL}=Q_{\mu\nu}^{C}\left(1-\frac{\gamma}{\sqrt{N_{\mu\nu}Q_{\mu\nu}^{C}}}\right).
\end{aligned}
\end{equation}
where $\gamma$ is the number of standard deviations, and
\begin{equation} \label{fp3}
\begin{aligned}
\epsilon'= \frac{1}{\sqrt{2\pi}}\int_{\gamma}^{\infty}e^{-\frac{t^2}{2}}dt,
\end{aligned}
\end{equation}
where $\epsilon'$ is the failure probability for each estimation.

The data string of $\mu0$, $0\mu$, $\nu\nu$, $\nu0$, $0\nu$ and $00$ are all announced publicly to estimate the bit error rate of single-photon pairs in Eq. \eqref{error3}.
Therefore, the data string under the case of three intensity sets $\mu\mu$, $\mu\nu$ and $\nu\mu$ constitutes an overall data string $M$, i.e., $M=M_{\mu\mu}+M_{\mu\nu}+M_{\nu\mu}$.
Similarly, $M^{C}=M_{\mu\mu}^{C}+M_{\mu\nu}^{C}+M_{\nu\mu}^{C}$ and $M^{B}=M_{\mu\mu}^{B}+M_{\mu\nu}^{B}+M_{\nu\mu}^{B}$ are the numbers of Bob's and Charlie's conclusive results in the overall data string $M$, respectively.
We denote $M_{s}$ to be the sampling data string, $M_{r}=M-M_{s}$ to be the rest data string, $M_{s}^{B}$ and $M_{s}^{C}$ to be the numbers of Bob's and Charlie's conclusive results in the sampling data string $M_{s}$, $M_{r}^{B}=M^{B}-M_{s}^{B}$ and $M_{r}^{C}=M^{C}-M_{s}^{C}$ to be the numbers of Bob's and Charlie's conclusive results in the rest data string $M_{t}$. We denote $E_{s}^{B}$ and $E_{s}^{C}$ to be the quantum bit error rates in $M_{s}^B$ and $M_{s}^{C}$, respectively.

From the view of Bob and Charlie, only the conclusive results can be used to detect the error (mismatching), while the inconclusive results can only be assumed without mismatching.
The mismatching rates of Bob's and Charlie's in the sampling data string can be given by
\begin{equation} \label{dis}
\begin{aligned}
\Delta_{s}^{B}=E_{s}^{B}M_{s}^{B}/M_{s},~~~   \Delta_{s}^{C}=E_{s}^{C}M_{s}^{C}/M_{s}.
\end{aligned}
\end{equation}
However, from the view of Alice, the status of all data owned by Bob (Charlie) in the data string $M$ could be regarded as the same, since Bob (Charlie) does not announce the position of conclusive results.
By using the random sampling without replacement \cite{korzh2015provably}, the upper bound of the difference between the data owned by Bob and by Charlie in the rest data string can be given by
\begin{equation} \label{dis}
\begin{aligned}
\Delta=\Delta_{s}+\delta,~~\Delta_{s}=\Delta_{s}^{B}+\Delta_{s}^{C},~~\delta=g[M_{s},M_{r},\Delta_{s},\epsilon],
\end{aligned}
\end{equation}
where $\epsilon=10^{-6}$ is the failure probability and
\begin{equation}
\begin{aligned}
g(n,k,\lambda,\epsilon)=&\sqrt{\frac{2(n+k)\lambda(1-\lambda)}{nk}\ln\frac{\sqrt{n+k}C(n,k,\lambda)}{\sqrt{2\pi nk\lambda(1-\lambda)}\epsilon}},\\
C(n,k,\lambda)=&\textrm{exp}\Big(\frac{1}{8(n+k)}+\frac{1}{12k}-\frac{1}{12k\lambda+1}-\frac{1}{12k(1-\lambda)+1}\Big).
\end{aligned}
\end{equation}

By using the Chernoff Bound \cite{chernoff1952measure,curty2014finite}, the optimal probability of Alice's repudiation attack can be written as \cite{Yin:2016:Practical}
\begin{equation} \label{pro}
\begin{aligned}
\varepsilon_{\textrm{rep}}=\exp\left[-\frac{(A-M_{r}^{B}T_{a}/M_{r})^2}{2A}M_{r}\right],\\
\end{aligned}
\end{equation}
where $A$ is the physical solution of the following equation and inequalities,
\begin{equation} \label{sol}
\begin{aligned}
&\frac{(A-M_{r}^{B}T_{a}/M_{r})^2}{2A}=\frac{\left[M_{r}^{C}T_{v}/M_{r}-(A+\Delta)\right]^2}{3(A+\Delta)},\\
&M_{r}^{B}T_{a}/M_{r}<A<(M_{r}^{C}T_{v}/M_{r}-\Delta).
\end{aligned}
\end{equation}

The number of the single-photon pairs of Charlie's conclusive results in the rest data string can be given by
\begin{equation}
\begin{aligned}
M_{11r}^{C}=(M_{r}/M)(N_{\mu\mu}e^{-2\mu}\mu^2+N_{\mu\nu}e^{-\mu-\nu}\mu\nu+N_{\nu\mu}e^{-\mu-\nu}\mu\nu)Y_{11}^{C}.
\end{aligned}
\end{equation}
We assume that Bob can guess the information of Charlie's conclusive results without error unless the single-photon pairs. The lower bound of mismatching rate $S_{11}$ between Bob's declaration and Charlie's conclusive results with single-photon pairs can be given by
\begin{equation} \label{qubit}
\begin{aligned}
1-H(e_{p11}^{C}|e_{11}^{C})-H(S_{11})=0,
\end{aligned}
\end{equation}
where $H(x)=-x\log_{2}x-(1-x)\log_{2}(1-x)$ is the Shannon entropy function, the phase error rate $e_{p11}^{C}$ of single-photon pairs can be calculated by Eq. \eqref{eq1} and $a=\frac{2-\sqrt{2}}{4}$.
The optimal probability of Bob's forgery attack is \cite{Yin:2016:Practical}
\begin{equation} \label{pro}
\begin{aligned}
\varepsilon_{\textrm{for}}=\exp\left[-\frac{(S_{11}-T_{v11})^2}{2S_{11}}M_{r11}^{C}\right],
\end{aligned}
\end{equation}
where $T_{v11}=T_{v}M_{r}^{C}/M_{r11}^{C}$ is the error rate threshold of single-photon pairs of Charlie's conclusive results.
The secure bound of the protocol can be written as
\begin{equation} \label{sol}
\begin{aligned}
\varepsilon_{\textrm{sec}}&=\varepsilon_{\textrm{for}}+\varepsilon_{\textrm{rep}}+\epsilon+11\epsilon'.
\end{aligned}
\end{equation}
where $11\epsilon'=7\times 10^{-6}$ is the failure probability due to the decoy-state method.

The probability of the robustness is \cite{Yin:2016:Practical}
\begin{equation} \label{rob}
\begin{aligned}
\varepsilon_{\textrm{rob}}&=h[M_{r},M_{s},\Delta_{s}^{B}, M_{r}^{B}T_{a}/M_{r}-\Delta_{s}^{B}].
\end{aligned}
\end{equation}
where
\begin{equation} \label{rob}
\begin{aligned}
h(n,k,\lambda,t)=&\frac{\exp[-\frac{nkt^{2}}{2(n+k)\lambda(1-\lambda)}]C(n,k,\lambda)}{\sqrt{2\pi nk\lambda(1-\lambda)/(n+k)}}.
\end{aligned}
\end{equation}

\section{Experimental results}

We have performed the QDS experiment in the laboratory. The distances form Alice to Bob (Alice to Charlie) are performed with four cases, i.e., 25 km, 51 km, 76 km and 102 km fiber spools.
Therefore, the maximum distances between Bob and Charlie can be about 50 km, 102 km, 152 km, 204 km. The secure parameters and important results at different distances in the experiment are shown in Table \ref{tab:Key}. Tables \ref{tab:count1}-\ref{tab:count4} show the details of the total pulses $N_{\alpha\beta}$, the total counts $M_{\alpha\beta}$, $M_{\alpha\beta}^{B}$, $M_{\alpha\beta}^{C}$, the error rates $E_{\alpha\beta}^{B}$ and $E_{\alpha\beta}^{C}$. From the experimental results, we can see that the probabilities of Bob's and Charlie's conclusive results are all approximately 0.25 and in accordance with the theory.
Table \ref{tab:count5} shows the case of the random sampling with the probability of 30\%.

\begin{table}
\begin{ruledtabular}
\caption{\label{tab:Key} The secure parameters at different distances in the experiment.  }
\begin{tabular}{c|cccccccc}
Distance
&
\multicolumn{2}{c}{25km}
&
\multicolumn{2}{c}{51km}
&
\multicolumn{2}{c}{76km}
&
\multicolumn{2}{c}{102km}\\

Attenuation
&
\multicolumn{2}{c}{4.9dB}
&
\multicolumn{2}{c}{9.8dB}
&
\multicolumn{2}{c}{14.8dB}
&
\multicolumn{2}{c}{19.8dB}\\

$T_{v}$
&
\multicolumn{2}{c}{2.0\%}
&
\multicolumn{2}{c}{2.0\%}
&
\multicolumn{2}{c}{1.9\%}
&
\multicolumn{2}{c}{2.2\%}\\

$T_{a}$
&
\multicolumn{2}{c}{0.6\%}
&
\multicolumn{2}{c}{0.6\%}
&
\multicolumn{2}{c}{0.55\%}
&
\multicolumn{2}{c}{0.7\%}\\
\hline
Message
& m=0 & m=1 & m=0 & m=1 & m=0 & m=1 & m=0 & m=1\\
Total pulse
& $1.5\times10^9$ & $1.5\times10^9$ & $1.35\times10^{10}$ & $1.35\times10^{10}$ & $1.215\times10^{11}$ & $1.215\times10^{11}$ & $2.5065\times10^{12}$ & $2.5065\times10^{12}$\\
Time
& 20s & 20s & 180s & 180s & 1620s & 1620s & 33420s & 33420s \\
$e_{11}^{C}$
& 1.12\%& 1.20\% & 1.16\% & 1.10\% & 1.15\% & 1.29\% & 1.03\% & 1.05\%\\
$Y_{11}^{C}$
& $2.74\times10^{-3}$ & $2.55\times10^{-3}$ & $3.20\times10^{-4}$ & $3.09\times10^{-4}$ & $3.29\times10^{-5}$ & $3.42\times10^{-5}$ & $3.42\times10^{-6}$ & $3.48\times10^{-6}$\\
$S_{11}$
& 4.33\%& 4.21\% & 4.27\% & 4.35\% & 4.28\% & 4.10\% & 4.46\% & 4.42\%\\
$\varepsilon_{\textrm{rep}}$
& $7.1\times10^{-10}$ & $1.4\times10^{-6}$ & $1.6\times10^{-10}$ & $3.4\times10^{-11}$ & $5.3\times10^{-8}$ & $4.9\times10^{-12}$ & $4.9\times10^{-8}$ & $5.7\times10^{-13}$\\
$\varepsilon_{\textrm{for}}$
& $7.4\times10^{-15}$ & $5.6\times10^{-9}$ & $4.1\times10^{-13}$ & $4.1\times10^{-12}$ & $2.5\times10^{-8}$ & $3.7\times10^{-9}$ & $1.4\times10^{-19}$ & $7.0\times10^{-10}$\\
$\varepsilon_{\textrm{rob}}$
& $8.2\times10^{-12}$ & $3.9\times10^{-8}$ & $1.6\times10^{-9}$ & $4.6\times10^{-10}$ & $1.2\times10^{-7}$ & $1.2\times10^{-14}$ & $2.2\times10^{-9}$ & $2.0\times10^{-18}$\\
\end{tabular}
\end{ruledtabular}
\end{table}

\begin{table}
\centering
\caption{List of the total pulses, the total counts and the error counts in the case of 25 km in the laboratory.} \label{tab:count1}
\begin{tabular}{c|c|ccc|ccc}
\hline
\hline
$25 km$ & $$ & &m=0&   & &m=1   \\
\hline
$$ & & $0$&$\nu$&$\mu$ & $0$&$\nu$&$\mu$ \\
\hline
$$ &     $0$ &        4.13E+06	&2.62E+07	&4.47E+07& 4.13E+06	&2.62E+07	&4.47E+07\\
$N_{\mu\nu}$&$\nu$ &  2.64E+07	&1.85E+08	&3.14E+08& 2.64E+07	&1.85E+08	&3.14E+08\\
$$ &     $\mu$ &     4.45E+07	&3.14E+08	&5.42E+08	&4.45E+07	&3.14E+08	&5.42E+08\\
\hline
$$ &     $0$        & 1	&4	&13& 0	&2	&13  \\
$M_{\mu\nu}$&$\nu$  & 3	&10743	&59590& 2	&10206	&56841  \\
$$ &     $\mu$      & 4	&59339	&340524& 9	&56775	&323515  \\
\hline
$$ &     $0$           & 0	&0	&7  & 0	&0	&6  \\
$M_{\mu\nu}^{B}$&$\nu$ & 1	&2715	&14937  &1	&2545	&14330  \\
$$ &     $\mu$         & 3	&14778	&85049  & 0	&14320	&81006  \\
\hline
$$ &     $0$           & 0	&1	&4  & 0	&1	&6   \\
$M_{\mu\nu}^{C}$&$\nu$ & 1	&2768	&14839  & 1	&2614	&14215 \\
$$ &     $\mu$         & 2	&14990	&84844  &  4	&14108	&80513  \\
\hline
$$ &     $0$           & 0	&0	&5  & 0	&0	&4  \\
$E_{\mu\nu}^{B}M_{\mu\nu}^{B}$&$\nu$ & 0	&13	&61   & 0	&10	&83  \\
$$ &     $\mu$         & 0	&52	&283    & 0	&48	&335  \\
\hline
$$ &     $0$           &0	&0	&0  & 0	&0	&0  \\
$E_{\mu\nu}^{C}M_{\mu\nu}^{C}$&$\nu$ & 1	&8	&31  & 1	&8	&43  \\
$$ &     $\mu$         & 1	&58	&188  & 3	&53	&212  \\
\hline
\hline
\end{tabular}
\end{table}

\begin{table}
\centering
\caption{List of the total pulses, the total counts and the error counts in the case of 51 km in the laboratory.} \label{tab:count2}
\begin{tabular}{c|c|ccc|ccc}
\hline
\hline
$51 km$ & $$ & &m=0&   & &m=1   \\
\hline
$$ & & $0$&$\nu$&$\mu$ & $0$&$\nu$&$\mu$ \\
\hline
$$ &     $0$ &        3.72E+07	&2.35E+08	&4.02E+08& 3.72E+07	&2.35E+08	&4.02E+08\\
$N_{\mu\nu}$&$\nu$ &  2.37E+08	&1.66E+09	&2.82E+09& 2.37E+08	&1.66E+09	&2.82E+09\\
$$ &     $\mu$ &     4.01E+08	&2.82E+09	&4.87E+09	&4.01E+08	&2.82E+09	&4.87E+09\\
\hline
$$ &     $0$        & 0	&2	&4& 0	&0	&5  \\
$M_{\mu\nu}$&$\nu$  & 0	&10958	&62904& 1	&11334	&62323  \\
$$ &     $\mu$      & 2	&61726	&359788& 2	&61619	&356586  \\
\hline
$$ &     $0$           & 0	&0	&2  & 0	&0	&2  \\
$M_{\mu\nu}^{B}$&$\nu$ & 0	&2805	&15806  &0	&2923	&15695  \\
$$ &     $\mu$         & 2	&15312	&89669  & 0	&15506	&89527  \\
\hline
$$ &     $0$           & 0	&0	&0  & 0&	0	&0   \\
$M_{\mu\nu}^{C}$&$\nu$ &0	&2835	&15877  & 1	&2818	&15491\\
$$ &     $\mu$         & 0	&15518	&90300  &  0	&15382	&89748  \\
\hline
$$ &     $0$           & 0	&0	&1  & 0&	0	&1  \\
$E_{\mu\nu}^{B}M_{\mu\nu}^{B}$&$\nu$ & 0	&12	&57  & 0	&15	&54  \\
$$ &     $\mu$         & 0	&70	&372    & 0	&59	&294  \\
\hline
$$ &     $0$           &0	&0	&0  & 0	&0	&0  \\
$E_{\mu\nu}^{C}M_{\mu\nu}^{C}$&$\nu$ & 0	&9	&44  & 0	&8	&30  \\
$$ &     $\mu$         & 0	&40	&204  &0	&31	&192  \\
\hline
\hline
\end{tabular}
\end{table}

\begin{table}
\centering
\caption{List of the total pulses, the total counts and the error counts in the case of 76 km in the laboratory.} \label{tab:count3}
\begin{tabular}{c|c|ccc|ccc}
\hline
\hline
$76 km$ & $$ & &m=0&   & &m=1   \\
\hline
$$ & & $0$&$\nu$&$\mu$ & $0$&$\nu$&$\mu$ \\
\hline
$$ &     $0$ &        3.34E+08	&2.12E+09	&3.62E+09 &3.34E+08	&2.12E+09	&3.62E+09\\
$N_{\mu\nu}$&$\nu$ &  2.13E+09	&1.50E+10	&2.54E+10&2.13E+09	&1.50E+10	&2.54E+10\\
$$ &     $\mu$ &     3.60E+09	&2.54E+10	&4.39E+10	&3.60E+09	&2.54E+10	&4.39E+10\\
\hline
$$ &     $0$        & 0	&0		&4 & 0	&0	&6  \\
$M_{\mu\nu}$&$\nu$  & 2	&10912	&62070& 1	&10794	&61148  \\
$$ &     $\mu$      & 6	&62252	&363460& 3	&62362	&360661  \\
\hline
$$ &     $0$           & 0	&0	&2  & 0	&0	&0  \\
$M_{\mu\nu}^{B}$&$\nu$ & 1	&2681	&15491  &1	&2722	&15346  \\
$$ &     $\mu$         & 2	&15781	&91014  & 0	&15670	&90341  \\
\hline
$$ &     $0$           &0	&0	&3  & 0	&0	&1   \\
$M_{\mu\nu}^{C}$&$\nu$ &0	&2741	&15779  & 0	&2785	&15198\\
$$ &     $\mu$         &2	&15716	&91624  &  0	&15543	&90804  \\
\hline
$$ &     $0$           & 0	&0	&1  & 0	&0	&0  \\
$E_{\mu\nu}^{B}M_{\mu\nu}^{B}$&$\nu$ & 0	&14	&50  & 0	&12&	44 \\
$$ &     $\mu$         & 0	&52	&297    & 0	&45	&261  \\
\hline
$$ &     $0$           &0	&0	&0  & 0	&0	&0  \\
$E_{\mu\nu}^{C}M_{\mu\nu}^{C}$&$\nu$ & 0	&8	&63  & 0	&10 &35  \\
$$ &     $\mu$         & 1	&45	&272  &0	&60	&265  \\
\hline
\hline
\end{tabular}
\end{table}

\begin{table}
\centering
\caption{List of the total pulses, the total counts and the error counts in the case of 102 km in the laboratory.} \label{tab:count4}
\begin{tabular}{c|c|ccc|ccc}
\hline
\hline
$102 km$ & $$ & &m=0&   & &m=1   \\
\hline
$$ & & $0$&$\nu$&$\mu$ & $0$&$\nu$&$\mu$ \\
\hline
$$ &     $0$ &        6.90E+09	&4.37E+10	&7.47E+10 &6.90E+09	&4.37E+10	&7.47E+10\\
$N_{\mu\nu}$&$\nu$ &  4.41E+10	&3.09E+11	&5.24E+11 &4.41E+10	&3.09E+11	&5.24E+11\\
$$ &     $\mu$ &     7.44E+10	&5.24E+11	&9.05E+11	&7.44E+10	&5.24E+11	&9.05E+11\\
\hline
$$ &     $0$        & 0	&1	&13 & 0&	3	&8  \\
$M_{\mu\nu}$&$\nu$  & 1	&21333	&120069& 1	&22744	&127079  \\
$$ &     $\mu$      & 11	&120257	&698071& 6	&127284	&740794  \\
\hline
$$ &     $0$           & 0	&0	&8  & 0&	0	&2  \\
$M_{\mu\nu}^{B}$&$\nu$ & 0	&5371	&30050  &1	&5640	&31713  \\
$$ &     $\mu$         & 2	&29996	&175595  & 3	&32088	&186024  \\
\hline
$$ &     $0$           &0	&0	&2  &0	&1	&1   \\
$M_{\mu\nu}^{C}$&$\nu$ &0	&5411	&30174  & 1	&5668	&31701\\
$$ &     $\mu$         &5	&30267	&175246  &  3	&32217	&185762  \\
\hline
$$ &     $0$           & 0	&0	&2  & 0	&0	&0  \\
$E_{\mu\nu}^{B}M_{\mu\nu}^{B}$&$\nu$ & 0	&31	&182  & 0	&34&	127 \\
$$ &     $\mu$         & 0	&154	&856    & 1	&145	&767  \\
\hline
$$ &     $0$           &0	&0	&0  & 0	&0	&0  \\
$E_{\mu\nu}^{C}M_{\mu\nu}^{C}$&$\nu$ & 0	&20	&113  & 0	&21	&162 \\
$$ &     $\mu$         & 2	&147	&799  &0	&140	&820  \\
\hline
\hline
\end{tabular}
\end{table}

\begin{table}
\begin{ruledtabular}
\caption{\label{tab:count5} The counts and error rates of the random sampling.  }
\begin{tabular}{c|cccccccc}
Distance
&
\multicolumn{2}{c}{25km}
&
\multicolumn{2}{c}{51km}
&
\multicolumn{2}{c}{76km}
&
\multicolumn{2}{c}{102km}\\
\hline
Message
& m=0 & m=1 & m=0 & m=1 & m=0 & m=1 & m=0 & m=1\\
$M_{s}$
& 137597 & 131162 & 145683 & 144388 & 146108 & 145195 & 281435 & 298206\\
$M_{r}$
& 321856 & 305969& 338735 & 336140 & 341674 & 338976 & 656962 & 696951 \\
$M_{s}^{B}$
& 34378& 32689 & 36542 & 36247 & 36650 & 36634 & 70967 & 74697\\
$M_{r}^{B}$
& 80386 & 76967 & 84245 & 84481 & 85636 & 84723 & 164674 & 175128\\
$M_{s}^{C}$
& 34203 & 32660 & 36491 & 36226 & 36754 & 36529 & 70616 & 74828\\
$M_{r}^{C}$
& 80470& 76176 & 85204 & 84395 & 86365 & 85016 & 165071 & 174852\\
$E_{s}^{B}M_{s}^{B}$
& 119 &127 & 136 & 134 & 132 & 106 & 364	 & 336\\
$E_{s}^{C}M_{s}^{C}$
& 90 & 95 & 90 & 80 & 110 & 94 & 297 & 300\\
\end{tabular}
\end{ruledtabular}
\end{table}



\begin{thebibliography}{32}%
\makeatletter
\providecommand \@ifxundefined [1]{%
 \@ifx{#1\undefined}
}%
\providecommand \@ifnum [1]{%
 \ifnum #1\expandafter \@firstoftwo
 \else \expandafter \@secondoftwo
 \fi
}%
\providecommand \@ifx [1]{%
 \ifx #1\expandafter \@firstoftwo
 \else \expandafter \@secondoftwo
 \fi
}%
\providecommand \natexlab [1]{#1}%
\providecommand \enquote  [1]{``#1''}%
\providecommand \bibnamefont  [1]{#1}%
\providecommand \bibfnamefont [1]{#1}%
\providecommand \citenamefont [1]{#1}%
\providecommand \href@noop [0]{\@secondoftwo}%
\providecommand \href [0]{\begingroup \@sanitize@url \@href}%
\providecommand \@href[1]{\@@startlink{#1}\@@href}%
\providecommand \@@href[1]{\endgroup#1\@@endlink}%
\providecommand \@sanitize@url [0]{\catcode `\\12\catcode `\$12\catcode
  `\&12\catcode `\#12\catcode `\^12\catcode `\_12\catcode `\%12\relax}%
\providecommand \@@startlink[1]{}%
\providecommand \@@endlink[0]{}%
\providecommand \url  [0]{\begingroup\@sanitize@url \@url }%
\providecommand \@url [1]{\endgroup\@href {#1}{\urlprefix }}%
\providecommand \urlprefix  [0]{URL }%
\providecommand \Eprint [0]{\href }%
\providecommand \doibase [0]{http://dx.doi.org/}%
\providecommand \selectlanguage [0]{\@gobble}%
\providecommand \bibinfo  [0]{\@secondoftwo}%
\providecommand \bibfield  [0]{\@secondoftwo}%
\providecommand \translation [1]{[#1]}%
\providecommand \BibitemOpen [0]{}%
\providecommand \bibitemStop [0]{}%
\providecommand \bibitemNoStop [0]{.\EOS\space}%
\providecommand \EOS [0]{\spacefactor3000\relax}%
\providecommand \BibitemShut  [1]{\csname bibitem#1\endcsname}%
\let\auto@bib@innerbib\@empty
\bibitem [{\citenamefont {Diffie}\ and\ \citenamefont
  {Hellman}(1976)}]{diffie:1976:new}%
  \BibitemOpen
  \bibfield  {author} {\bibinfo {author} {\bibfnamefont {W.}~\bibnamefont
  {Diffie}}\ and\ \bibinfo {author} {\bibfnamefont {M.}~\bibnamefont
  {Hellman}},\ }\href@noop {} {\bibfield  {journal} {\bibinfo  {journal} {IEEE
  Transactions on Information Theory}\ }\textbf {\bibinfo {volume} {22}},\
  \bibinfo {pages} {644} (\bibinfo {year} {1976})}\BibitemShut {NoStop}%
\bibitem [{\citenamefont {Rivest}\ \emph {et~al.}(1978)\citenamefont {Rivest},
  \citenamefont {Shamir},\ and\ \citenamefont {Adleman}}]{rivest:1978:method}%
  \BibitemOpen
  \bibfield  {author} {\bibinfo {author} {\bibfnamefont {R.~L.}\ \bibnamefont
  {Rivest}}, \bibinfo {author} {\bibfnamefont {A.}~\bibnamefont {Shamir}}, \
  and\ \bibinfo {author} {\bibfnamefont {L.}~\bibnamefont {Adleman}},\
  }\href@noop {} {\bibfield  {journal} {\bibinfo  {journal} {Communications of
  the ACM}\ }\textbf {\bibinfo {volume} {21}},\ \bibinfo {pages} {120}
  (\bibinfo {year} {1978})}\BibitemShut {NoStop}%
\bibitem [{\citenamefont {Shor}(1994)}]{shor:1994:algorithms}%
  \BibitemOpen
  \bibfield  {author} {\bibinfo {author} {\bibfnamefont {P.~W.}\ \bibnamefont
  {Shor}},\ }in\ \href@noop {} {\emph {\bibinfo {booktitle} {Foundations of
  Computer Science, 1994 Proceedings., 35th Annual Symposium on}}}\ (\bibinfo
  {organization} {IEEE},\ \bibinfo {year} {1994})\ pp.\ \bibinfo {pages}
  {124--134}\BibitemShut {NoStop}%
\bibitem [{\citenamefont {Gisin}\ \emph {et~al.}(2002)\citenamefont {Gisin},
  \citenamefont {Ribordy}, \citenamefont {Tittel},\ and\ \citenamefont
  {Zbinden}}]{Gisin:2002:Quantum}%
  \BibitemOpen
  \bibfield  {author} {\bibinfo {author} {\bibfnamefont {N.}~\bibnamefont
  {Gisin}}, \bibinfo {author} {\bibfnamefont {G.}~\bibnamefont {Ribordy}},
  \bibinfo {author} {\bibfnamefont {W.}~\bibnamefont {Tittel}}, \ and\ \bibinfo
  {author} {\bibfnamefont {H.}~\bibnamefont {Zbinden}},\ }\href@noop {}
  {\bibfield  {journal} {\bibinfo  {journal} {Rev. Mod. Phys.}\ }\textbf
  {\bibinfo {volume} {74}},\ \bibinfo {pages} {145} (\bibinfo {year}
  {2002})}\BibitemShut {NoStop}%
\bibitem [{\citenamefont {Scarani}\ \emph {et~al.}(2009)\citenamefont
  {Scarani}, \citenamefont {Bechmann-Pasquinucci}, \citenamefont {Cerf},
  \citenamefont {Du\ifmmode~\check{s}\else \v{s}\fi{}ek}, \citenamefont
  {L\"utkenhaus},\ and\ \citenamefont {Peev}}]{Scarani:2009:The}%
  \BibitemOpen
  \bibfield  {author} {\bibinfo {author} {\bibfnamefont {V.}~\bibnamefont
  {Scarani}}, \bibinfo {author} {\bibfnamefont {H.}~\bibnamefont
  {Bechmann-Pasquinucci}}, \bibinfo {author} {\bibfnamefont {N.~J.}\
  \bibnamefont {Cerf}}, \bibinfo {author} {\bibfnamefont {M.}~\bibnamefont
  {Du\ifmmode~\check{s}\else \v{s}\fi{}ek}}, \bibinfo {author} {\bibfnamefont
  {N.}~\bibnamefont {L\"utkenhaus}}, \ and\ \bibinfo {author} {\bibfnamefont
  {M.}~\bibnamefont {Peev}},\ }\href@noop {} {\bibfield  {journal} {\bibinfo
  {journal} {Rev. Mod. Phys.}\ }\textbf {\bibinfo {volume} {81}},\ \bibinfo
  {pages} {1301} (\bibinfo {year} {2009})}\BibitemShut {NoStop}%
\bibitem [{\citenamefont {Qiu}(2014)}]{qiu:2014:quantum}%
  \BibitemOpen
  \bibfield  {author} {\bibinfo {author} {\bibfnamefont {J.}~\bibnamefont
  {Qiu}},\ }\href@noop {} {\bibfield  {journal} {\bibinfo  {journal} {Nature}\
  }\textbf {\bibinfo {volume} {508}},\ \bibinfo {pages} {441} (\bibinfo {year}
  {2014})}\BibitemShut {NoStop}%
\bibitem [{\citenamefont {Gottesman}\ and\ \citenamefont
  {Chuang}(2001)}]{gottesman:2001:quantum}%
  \BibitemOpen
  \bibfield  {author} {\bibinfo {author} {\bibfnamefont {D.}~\bibnamefont
  {Gottesman}}\ and\ \bibinfo {author} {\bibfnamefont {I.}~\bibnamefont
  {Chuang}},\ }\href@noop {} {\bibfield  {journal} {\bibinfo  {journal} {arXiv
  quant-ph/0105032}\ } (\bibinfo {year} {2001})}\BibitemShut {NoStop}%
\bibitem [{\citenamefont {Andersson}\ \emph {et~al.}(2006)\citenamefont
  {Andersson}, \citenamefont {Curty},\ and\ \citenamefont
  {Jex}}]{Andersson:2007:Experimentally}%
  \BibitemOpen
  \bibfield  {author} {\bibinfo {author} {\bibfnamefont {E.}~\bibnamefont
  {Andersson}}, \bibinfo {author} {\bibfnamefont {M.}~\bibnamefont {Curty}}, \
  and\ \bibinfo {author} {\bibfnamefont {I.}~\bibnamefont {Jex}},\ }\href@noop
  {} {\bibfield  {journal} {\bibinfo  {journal} {Phys. Rev. A}\ }\textbf
  {\bibinfo {volume} {74}},\ \bibinfo {pages} {022304} (\bibinfo {year}
  {2006})}\BibitemShut {NoStop}%
\bibitem [{\citenamefont {Dunjko}\ \emph {et~al.}(2014)\citenamefont {Dunjko},
  \citenamefont {Wallden},\ and\ \citenamefont
  {Andersson}}]{Dunjko:2014:Quantum}%
  \BibitemOpen
  \bibfield  {author} {\bibinfo {author} {\bibfnamefont {V.}~\bibnamefont
  {Dunjko}}, \bibinfo {author} {\bibfnamefont {P.}~\bibnamefont {Wallden}}, \
  and\ \bibinfo {author} {\bibfnamefont {E.}~\bibnamefont {Andersson}},\
  }\href@noop {} {\bibfield  {journal} {\bibinfo  {journal} {Phys. Rev. Lett.}\
  }\textbf {\bibinfo {volume} {112}},\ \bibinfo {pages} {040502} (\bibinfo
  {year} {2014})}\BibitemShut {NoStop}%
\bibitem [{\citenamefont {Arrazola}\ and\ \citenamefont
  {L\"utkenhaus}(2014)}]{Arrazola:2014:Quantum}%
  \BibitemOpen
  \bibfield  {author} {\bibinfo {author} {\bibfnamefont {J.~M.}\ \bibnamefont
  {Arrazola}}\ and\ \bibinfo {author} {\bibfnamefont {N.}~\bibnamefont
  {L\"utkenhaus}},\ }\href@noop {} {\bibfield  {journal} {\bibinfo  {journal}
  {Phys. Rev. A}\ }\textbf {\bibinfo {volume} {90}},\ \bibinfo {pages} {042335}
  (\bibinfo {year} {2014})}\BibitemShut {NoStop}%
\bibitem [{\citenamefont {Wallden}\ \emph {et~al.}(2015)\citenamefont
  {Wallden}, \citenamefont {Dunjko}, \citenamefont {Kent},\ and\ \citenamefont
  {Andersson}}]{Wallden:2015:Quantum}%
  \BibitemOpen
  \bibfield  {author} {\bibinfo {author} {\bibfnamefont {P.}~\bibnamefont
  {Wallden}}, \bibinfo {author} {\bibfnamefont {V.}~\bibnamefont {Dunjko}},
  \bibinfo {author} {\bibfnamefont {A.}~\bibnamefont {Kent}}, \ and\ \bibinfo
  {author} {\bibfnamefont {E.}~\bibnamefont {Andersson}},\ }\href@noop {}
  {\bibfield  {journal} {\bibinfo  {journal} {Phys. Rev. A}\ }\textbf {\bibinfo
  {volume} {91}},\ \bibinfo {pages} {042304} (\bibinfo {year}
  {2015})}\BibitemShut {NoStop}%
\bibitem [{\citenamefont {Arrazola}\ \emph {et~al.}(2015)\citenamefont
  {Arrazola}, \citenamefont {Wallden},\ and\ \citenamefont
  {Andersson}}]{arrazola2015multiparty}%
  \BibitemOpen
  \bibfield  {author} {\bibinfo {author} {\bibfnamefont {J.~M.}\ \bibnamefont
  {Arrazola}}, \bibinfo {author} {\bibfnamefont {P.}~\bibnamefont {Wallden}}, \
  and\ \bibinfo {author} {\bibfnamefont {E.}~\bibnamefont {Andersson}},\
  }\href@noop {} {\bibfield  {journal} {\bibinfo  {journal} {Quantum Inf.
  Comput.}\ }\textbf {\bibinfo {volume} {6}},\ \bibinfo {pages} {0435}
  (\bibinfo {year} {2015})}\BibitemShut {NoStop}%
\bibitem [{\citenamefont {Clarke}\ \emph {et~al.}(2012)\citenamefont {Clarke},
  \citenamefont {Collins}, \citenamefont {Dunjko}, \citenamefont {Andersson},
  \citenamefont {Jeffers},\ and\ \citenamefont
  {Buller}}]{clarke2012experimental}%
  \BibitemOpen
  \bibfield  {author} {\bibinfo {author} {\bibfnamefont {P.~J.}\ \bibnamefont
  {Clarke}}, \bibinfo {author} {\bibfnamefont {R.~J.}\ \bibnamefont {Collins}},
  \bibinfo {author} {\bibfnamefont {V.}~\bibnamefont {Dunjko}}, \bibinfo
  {author} {\bibfnamefont {E.}~\bibnamefont {Andersson}}, \bibinfo {author}
  {\bibfnamefont {J.}~\bibnamefont {Jeffers}}, \ and\ \bibinfo {author}
  {\bibfnamefont {G.~S.}\ \bibnamefont {Buller}},\ }\href@noop {} {\bibfield
  {journal} {\bibinfo  {journal} {Nature Commun.}\ }\textbf {\bibinfo {volume}
  {3}},\ \bibinfo {pages} {1174} (\bibinfo {year} {2012})}\BibitemShut
  {NoStop}%
\bibitem [{\citenamefont {Collins}\ \emph {et~al.}(2014)\citenamefont
  {Collins}, \citenamefont {Donaldson}, \citenamefont {Dunjko}, \citenamefont
  {Wallden}, \citenamefont {Clarke}, \citenamefont {Andersson}, \citenamefont
  {Jeffers},\ and\ \citenamefont {Buller}}]{Collins:2014:Realization}%
  \BibitemOpen
  \bibfield  {author} {\bibinfo {author} {\bibfnamefont {R.~J.}\ \bibnamefont
  {Collins}}, \bibinfo {author} {\bibfnamefont {R.~J.}\ \bibnamefont
  {Donaldson}}, \bibinfo {author} {\bibfnamefont {V.}~\bibnamefont {Dunjko}},
  \bibinfo {author} {\bibfnamefont {P.}~\bibnamefont {Wallden}}, \bibinfo
  {author} {\bibfnamefont {P.~J.}\ \bibnamefont {Clarke}}, \bibinfo {author}
  {\bibfnamefont {E.}~\bibnamefont {Andersson}}, \bibinfo {author}
  {\bibfnamefont {J.}~\bibnamefont {Jeffers}}, \ and\ \bibinfo {author}
  {\bibfnamefont {G.~S.}\ \bibnamefont {Buller}},\ }\href@noop {} {\bibfield
  {journal} {\bibinfo  {journal} {Phys. Rev. Lett.}\ }\textbf {\bibinfo
  {volume} {113}},\ \bibinfo {pages} {040502} (\bibinfo {year}
  {2014})}\BibitemShut {NoStop}%
\bibitem [{\citenamefont {Donaldson}\ \emph {et~al.}(2016)\citenamefont
  {Donaldson}, \citenamefont {Collins}, \citenamefont {Kleczkowska},
  \citenamefont {Amiri}, \citenamefont {Wallden}, \citenamefont {Dunjko},
  \citenamefont {Jeffers}, \citenamefont {Andersson},\ and\ \citenamefont
  {Buller}}]{Donaldson:2016:Experimental}%
  \BibitemOpen
  \bibfield  {author} {\bibinfo {author} {\bibfnamefont {R.~J.}\ \bibnamefont
  {Donaldson}}, \bibinfo {author} {\bibfnamefont {R.~J.}\ \bibnamefont
  {Collins}}, \bibinfo {author} {\bibfnamefont {K.}~\bibnamefont
  {Kleczkowska}}, \bibinfo {author} {\bibfnamefont {R.}~\bibnamefont {Amiri}},
  \bibinfo {author} {\bibfnamefont {P.}~\bibnamefont {Wallden}}, \bibinfo
  {author} {\bibfnamefont {V.}~\bibnamefont {Dunjko}}, \bibinfo {author}
  {\bibfnamefont {J.}~\bibnamefont {Jeffers}}, \bibinfo {author} {\bibfnamefont
  {E.}~\bibnamefont {Andersson}}, \ and\ \bibinfo {author} {\bibfnamefont
  {G.~S.}\ \bibnamefont {Buller}},\ }\href@noop {} {\bibfield  {journal}
  {\bibinfo  {journal} {Phys. Rev. A}\ }\textbf {\bibinfo {volume} {93}},\
  \bibinfo {pages} {012329} (\bibinfo {year} {2016})}\BibitemShut {NoStop}%
\bibitem [{\citenamefont {Croal}\ \emph {et~al.}(2016)\citenamefont {Croal},
  \citenamefont {Peuntinger}, \citenamefont {Heim}, \citenamefont {Khan},
  \citenamefont {Marquardt}, \citenamefont {Leuchs}, \citenamefont {Wallden},
  \citenamefont {Andersson},\ and\ \citenamefont {Korolkova}}]{croal2016free}%
  \BibitemOpen
  \bibfield  {author} {\bibinfo {author} {\bibfnamefont {C.}~\bibnamefont
  {Croal}}, \bibinfo {author} {\bibfnamefont {C.}~\bibnamefont {Peuntinger}},
  \bibinfo {author} {\bibfnamefont {B.}~\bibnamefont {Heim}}, \bibinfo {author}
  {\bibfnamefont {I.}~\bibnamefont {Khan}}, \bibinfo {author} {\bibfnamefont
  {C.}~\bibnamefont {Marquardt}}, \bibinfo {author} {\bibfnamefont
  {G.}~\bibnamefont {Leuchs}}, \bibinfo {author} {\bibfnamefont
  {P.}~\bibnamefont {Wallden}}, \bibinfo {author} {\bibfnamefont
  {E.}~\bibnamefont {Andersson}}, \ and\ \bibinfo {author} {\bibfnamefont
  {N.}~\bibnamefont {Korolkova}},\ }\href@noop {} {\bibfield  {journal}
  {\bibinfo  {journal} {arXiv:1604.03708}\ } (\bibinfo {year}
  {2016})}\BibitemShut {NoStop}%
\bibitem [{\citenamefont {Hillery}\ \emph {et~al.}(1999)\citenamefont
  {Hillery}, \citenamefont {Bu{\v{z}}ek},\ and\ \citenamefont
  {Berthiaume}}]{hillery:1999:quantum}%
  \BibitemOpen
  \bibfield  {author} {\bibinfo {author} {\bibfnamefont {M.}~\bibnamefont
  {Hillery}}, \bibinfo {author} {\bibfnamefont {V.}~\bibnamefont
  {Bu{\v{z}}ek}}, \ and\ \bibinfo {author} {\bibfnamefont {A.}~\bibnamefont
  {Berthiaume}},\ }\href@noop {} {\bibfield  {journal} {\bibinfo  {journal}
  {Phy. Rev. A}\ }\textbf {\bibinfo {volume} {59}},\ \bibinfo {pages} {1829}
  (\bibinfo {year} {1999})}\BibitemShut {NoStop}%
\bibitem [{\citenamefont {Yin}\ \emph {et~al.}(2016{\natexlab{a}})\citenamefont
  {Yin}, \citenamefont {Fu},\ and\ \citenamefont {Chen}}]{Yin:2016:Practical}%
  \BibitemOpen
  \bibfield  {author} {\bibinfo {author} {\bibfnamefont {H.-L.}\ \bibnamefont
  {Yin}}, \bibinfo {author} {\bibfnamefont {Y.}~\bibnamefont {Fu}}, \ and\
  \bibinfo {author} {\bibfnamefont {Z.-B.}\ \bibnamefont {Chen}},\ }\href@noop
  {} {\bibfield  {journal} {\bibinfo  {journal} {Phys. Rev. A}\ }\textbf
  {\bibinfo {volume} {93}},\ \bibinfo {pages} {032316} (\bibinfo {year}
  {2016}{\natexlab{a}})}\BibitemShut {NoStop}%
\bibitem [{\citenamefont {Amiri}\ \emph {et~al.}(2016)\citenamefont {Amiri},
  \citenamefont {Wallden}, \citenamefont {Kent},\ and\ \citenamefont
  {Andersson}}]{Amiri:2016:Secure}%
  \BibitemOpen
  \bibfield  {author} {\bibinfo {author} {\bibfnamefont {R.}~\bibnamefont
  {Amiri}}, \bibinfo {author} {\bibfnamefont {P.}~\bibnamefont {Wallden}},
  \bibinfo {author} {\bibfnamefont {A.}~\bibnamefont {Kent}}, \ and\ \bibinfo
  {author} {\bibfnamefont {E.}~\bibnamefont {Andersson}},\ }\href@noop {}
  {\bibfield  {journal} {\bibinfo  {journal} {Phys. Rev. A}\ }\textbf {\bibinfo
  {volume} {93}},\ \bibinfo {pages} {032325} (\bibinfo {year}
  {2016})}\BibitemShut {NoStop}%
\bibitem [{\citenamefont {Bennett}\ and\ \citenamefont
  {Brassard}(1984)}]{bennett1984quantum}%
  \BibitemOpen
  \bibfield  {author} {\bibinfo {author} {\bibfnamefont {C.~H.}\ \bibnamefont
  {Bennett}}\ and\ \bibinfo {author} {\bibfnamefont {G.}~\bibnamefont
  {Brassard}},\ }in\ \href@noop {} {\emph {\bibinfo {booktitle} {International
  Conference on Computer System and Signal Processing, IEEE, 1984}}}\ (\bibinfo
  {year} {1984})\ pp.\ \bibinfo {pages} {175--179}\BibitemShut {NoStop}%
\bibitem [{\citenamefont {Scarani}\ \emph {et~al.}(2004)\citenamefont
  {Scarani}, \citenamefont {Ac\'{\i}n}, \citenamefont {Ribordy},\ and\
  \citenamefont {Gisin}}]{Scarani:2004:Quantum}%
  \BibitemOpen
  \bibfield  {author} {\bibinfo {author} {\bibfnamefont {V.}~\bibnamefont
  {Scarani}}, \bibinfo {author} {\bibfnamefont {A.}~\bibnamefont {Ac\'{\i}n}},
  \bibinfo {author} {\bibfnamefont {G.}~\bibnamefont {Ribordy}}, \ and\
  \bibinfo {author} {\bibfnamefont {N.}~\bibnamefont {Gisin}},\ }\href@noop {}
  {\bibfield  {journal} {\bibinfo  {journal} {Phys. Rev. Lett.}\ }\textbf
  {\bibinfo {volume} {92}},\ \bibinfo {pages} {057901} (\bibinfo {year}
  {2004})}\BibitemShut {NoStop}%
\bibitem [{\citenamefont {Tamaki}\ and\ \citenamefont
  {Lo}(2006)}]{Tamaki:2006:Unconditionally}%
  \BibitemOpen
  \bibfield  {author} {\bibinfo {author} {\bibfnamefont {K.}~\bibnamefont
  {Tamaki}}\ and\ \bibinfo {author} {\bibfnamefont {H.-K.}\ \bibnamefont
  {Lo}},\ }\href@noop {} {\bibfield  {journal} {\bibinfo  {journal} {Phys. Rev.
  A}\ }\textbf {\bibinfo {volume} {73}},\ \bibinfo {pages} {010302} (\bibinfo
  {year} {2006})}\BibitemShut {NoStop}%
\bibitem [{\citenamefont {Yin}\ \emph {et~al.}(2016{\natexlab{b}})\citenamefont
  {Yin}, \citenamefont {Fu}, \citenamefont {Mao},\ and\ \citenamefont
  {Chen}}]{Yin:2016:Security}%
  \BibitemOpen
  \bibfield  {author} {\bibinfo {author} {\bibfnamefont {H.-L.}\ \bibnamefont
  {Yin}}, \bibinfo {author} {\bibfnamefont {Y.}~\bibnamefont {Fu}}, \bibinfo
  {author} {\bibfnamefont {Y.}~\bibnamefont {Mao}}, \ and\ \bibinfo {author}
  {\bibfnamefont {Z.-B.}\ \bibnamefont {Chen}},\ }\href@noop {} {\bibfield
  {journal} {\bibinfo  {journal} {Sci. Rep.}\ }\textbf {\bibinfo {volume}
  {6}},\ \bibinfo {pages} {29482} (\bibinfo {year}
  {2016}{\natexlab{b}})}\BibitemShut {NoStop}%
\bibitem [{\citenamefont {Ma}\ \emph {et~al.}(2012)\citenamefont {Ma},
  \citenamefont {Fung},\ and\ \citenamefont {Razavi}}]{Ma:2012:Statistical}%
  \BibitemOpen
  \bibfield  {author} {\bibinfo {author} {\bibfnamefont {X.}~\bibnamefont
  {Ma}}, \bibinfo {author} {\bibfnamefont {C.-H.~F.}\ \bibnamefont {Fung}}, \
  and\ \bibinfo {author} {\bibfnamefont {M.}~\bibnamefont {Razavi}},\
  }\href@noop {} {\bibfield  {journal} {\bibinfo  {journal} {Phys. Rev. A}\
  }\textbf {\bibinfo {volume} {86}},\ \bibinfo {pages} {052305} (\bibinfo
  {year} {2012})}\BibitemShut {NoStop}%
\bibitem [{\citenamefont {Korzh}\ \emph {et~al.}(2015)\citenamefont {Korzh},
  \citenamefont {Lim}, \citenamefont {Houlmann}, \citenamefont {Gisin},
  \citenamefont {Li}, \citenamefont {Nolan}, \citenamefont {Sanguinetti},
  \citenamefont {Thew},\ and\ \citenamefont {Zbinden}}]{korzh2015provably}%
  \BibitemOpen
  \bibfield  {author} {\bibinfo {author} {\bibfnamefont {B.}~\bibnamefont
  {Korzh}}, \bibinfo {author} {\bibfnamefont {C.~C.~W.}\ \bibnamefont {Lim}},
  \bibinfo {author} {\bibfnamefont {R.}~\bibnamefont {Houlmann}}, \bibinfo
  {author} {\bibfnamefont {N.}~\bibnamefont {Gisin}}, \bibinfo {author}
  {\bibfnamefont {M.~J.}\ \bibnamefont {Li}}, \bibinfo {author} {\bibfnamefont
  {D.}~\bibnamefont {Nolan}}, \bibinfo {author} {\bibfnamefont
  {B.}~\bibnamefont {Sanguinetti}}, \bibinfo {author} {\bibfnamefont
  {R.}~\bibnamefont {Thew}}, \ and\ \bibinfo {author} {\bibfnamefont
  {H.}~\bibnamefont {Zbinden}},\ }\href@noop {} {\bibfield  {journal} {\bibinfo
   {journal} {Nature Photon.}\ }\textbf {\bibinfo {volume} {9}},\ \bibinfo
  {pages} {163} (\bibinfo {year} {2015})}\BibitemShut {NoStop}%
\bibitem [{\citenamefont {Chernoff}(1952)}]{chernoff1952measure}%
  \BibitemOpen
  \bibfield  {author} {\bibinfo {author} {\bibfnamefont {H.}~\bibnamefont
  {Chernoff}},\ }\href@noop {} {\bibfield  {journal} {\bibinfo  {journal} {Ann.
  Math. Stat.}\ }\textbf {\bibinfo {volume} {23}},\ \bibinfo {pages} {493}
  (\bibinfo {year} {1952})}\BibitemShut {NoStop}%
\bibitem [{\citenamefont {Curty}\ \emph {et~al.}(2014)\citenamefont {Curty},
  \citenamefont {Xu}, \citenamefont {Cui}, \citenamefont {Lim}, \citenamefont
  {Tamaki},\ and\ \citenamefont {Lo}}]{curty2014finite}%
  \BibitemOpen
  \bibfield  {author} {\bibinfo {author} {\bibfnamefont {M.}~\bibnamefont
  {Curty}}, \bibinfo {author} {\bibfnamefont {F.}~\bibnamefont {Xu}}, \bibinfo
  {author} {\bibfnamefont {W.}~\bibnamefont {Cui}}, \bibinfo {author}
  {\bibfnamefont {C.~C.~W.}\ \bibnamefont {Lim}}, \bibinfo {author}
  {\bibfnamefont {K.}~\bibnamefont {Tamaki}}, \ and\ \bibinfo {author}
  {\bibfnamefont {H.-K.}\ \bibnamefont {Lo}},\ }\href@noop {} {\bibfield
  {journal} {\bibinfo  {journal} {Nature Commun.}\ }\textbf {\bibinfo {volume}
  {5}},\ \bibinfo {pages} {3732} (\bibinfo {year} {2014})}\BibitemShut
  {NoStop}%
\bibitem [{\citenamefont {Hwang}(2003)}]{Hwang:2003:Decoy}%
  \BibitemOpen
  \bibfield  {author} {\bibinfo {author} {\bibfnamefont {W.-Y.}\ \bibnamefont
  {Hwang}},\ }\href@noop {} {\bibfield  {journal} {\bibinfo  {journal} {Phys.
  Rev. Lett.}\ }\textbf {\bibinfo {volume} {91}},\ \bibinfo {pages} {057901}
  (\bibinfo {year} {2003})}\BibitemShut {NoStop}%
\bibitem [{\citenamefont {Wang}(2005)}]{Wang:2005:Beating}%
  \BibitemOpen
  \bibfield  {author} {\bibinfo {author} {\bibfnamefont {X.-B.}\ \bibnamefont
  {Wang}},\ }\href@noop {} {\bibfield  {journal} {\bibinfo  {journal} {Phys.
  Rev. Lett.}\ }\textbf {\bibinfo {volume} {94}},\ \bibinfo {pages} {230503}
  (\bibinfo {year} {2005})}\BibitemShut {NoStop}%
\bibitem [{\citenamefont {Lo}\ \emph {et~al.}(2005)\citenamefont {Lo},
  \citenamefont {Ma},\ and\ \citenamefont {Chen}}]{Lo:2005:Decoy}%
  \BibitemOpen
  \bibfield  {author} {\bibinfo {author} {\bibfnamefont {H.-K.}\ \bibnamefont
  {Lo}}, \bibinfo {author} {\bibfnamefont {X.}~\bibnamefont {Ma}}, \ and\
  \bibinfo {author} {\bibfnamefont {K.}~\bibnamefont {Chen}},\ }\href@noop {}
  {\bibfield  {journal} {\bibinfo  {journal} {Phys. Rev. Lett.}\ }\textbf
  {\bibinfo {volume} {94}},\ \bibinfo {pages} {230504} (\bibinfo {year}
  {2005})}\BibitemShut {NoStop}%
\bibitem [{\citenamefont {Tang}\ \emph {et~al.}(2013)\citenamefont {Tang},
  \citenamefont {Yin}, \citenamefont {Ma}, \citenamefont {Fung}, \citenamefont
  {Liu}, \citenamefont {Yong}, \citenamefont {Chen}, \citenamefont {Peng},
  \citenamefont {Chen},\ and\ \citenamefont {Pan}}]{Tang:2013:USD}%
  \BibitemOpen
  \bibfield  {author} {\bibinfo {author} {\bibfnamefont {Y.-L.}\ \bibnamefont
  {Tang}}, \bibinfo {author} {\bibfnamefont {H.-L.}\ \bibnamefont {Yin}},
  \bibinfo {author} {\bibfnamefont {X.}~\bibnamefont {Ma}}, \bibinfo {author}
  {\bibfnamefont {C.-H.~F.}\ \bibnamefont {Fung}}, \bibinfo {author}
  {\bibfnamefont {Y.}~\bibnamefont {Liu}}, \bibinfo {author} {\bibfnamefont
  {H.-L.}\ \bibnamefont {Yong}}, \bibinfo {author} {\bibfnamefont {T.-Y.}\
  \bibnamefont {Chen}}, \bibinfo {author} {\bibfnamefont {C.-Z.}\ \bibnamefont
  {Peng}}, \bibinfo {author} {\bibfnamefont {Z.-B.}\ \bibnamefont {Chen}}, \
  and\ \bibinfo {author} {\bibfnamefont {J.-W.}\ \bibnamefont {Pan}},\
  }\href@noop {} {\bibfield  {journal} {\bibinfo  {journal} {Phys. Rev. A}\
  }\textbf {\bibinfo {volume} {88}},\ \bibinfo {pages} {022308} (\bibinfo
  {year} {2013})}\BibitemShut {NoStop}%
\bibitem [{\citenamefont {Zhou}\ \emph {et~al.}(2016)\citenamefont {Zhou},
  \citenamefont {Yu},\ and\ \citenamefont {Wang}}]{wang:2016:making}%
  \BibitemOpen
  \bibfield  {author} {\bibinfo {author} {\bibfnamefont {Y.-H.}\ \bibnamefont
  {Zhou}}, \bibinfo {author} {\bibfnamefont {Z.-W.}\ \bibnamefont {Yu}}, \ and\
  \bibinfo {author} {\bibfnamefont {X.-B.}\ \bibnamefont {Wang}},\ }\href@noop
  {} {\bibfield  {journal} {\bibinfo  {journal} {Phys. Rev. A}\ }\textbf
  {\bibinfo {volume} {93}},\ \bibinfo {pages} {042324} (\bibinfo {year}
  {2016})}\BibitemShut {NoStop}%
\end{thebibliography}

\end{document}